\documentclass[preprint2,tighten]{aastex}

\usepackage{xspace,color}

\singlespace

\bibpunct{(}{)}{;}{a}{}{,}

\shorttitle{The fossil group \RXJ1340}
\shortauthors{Mendes de Oliveira, Cypriano, Dupke \& Sodr\'e Jr.}

\def\sol{\mbox{$_{\odot}$}} 
\def\Msol{\hbox{M$_{\odot}$}}

\def\h50{\hbox{h$_{50}$}}
\def\m12{\hbox{$\Delta$m$_{12}$}}
\def\RXJ1340{RX~J1340.6+4018\xspace}
\def\1552{RX~J1552.2+2013\xspace}
\def\sloani{$i^\prime$\xspace}

\def\sloang{$g^\prime$\xspace}

\newbox\grsign \setbox\grsign=\hbox{$>$} \newdimen\grdimen
\grdimen=\ht\grsign
\newbox\simlessbox \newbox\simgreatbox
\setbox\simgreatbox=\hbox{\raise.5ex\hbox{$>$}\llap
    {\lower.5ex\hbox{$\sim$}}}\ht1=\grdimen\dp1=0pt
\setbox\simlessbox=\hbox{\raise.5ex\hbox{$<$}\llap
    {\lower.5ex\hbox{$\sim$}}}\ht2=\grdimen\dp2=0pt

\newbox\simppropto
\setbox\simppropto=\hbox{\raise.5ex\hbox{$\sim$}\llap
    {\lower.5ex\hbox{$\propto$}}}\ht2=\grdimen\dp2=0pt

\begin{document}

\title{
An optical and X-ray study of the fossil group \RXJ1340
\thanks{Based on observations obtained at the Gemini Observatory,
which is operated by the Association of Universities for Research in
Astronomy, Inc., under a cooperative agreement with the NSF on behalf of
the Gemini partnership: the National Science Foundation (United States),
the Particle Physics and Astronomy Research Council (United Kingdom),
the National Research Council (Canada), CONICYT (Chile), the Australian
Research Council (Australia), CNPq (Brazil) and CONICET (Argentina) --
Observing run ID: GN-2006B-Q-38.}}

\author{Claudia L. Mendes de Oliveira}
\affil{Departamento de Astronomia, Instituto de Astronomia, Geof\'{\i}sica
e Ci\^encias Atmosf\'ericas da USP, Rua do Mat\~ao 1226, Cidade
Universit\'aria, 05508-090, S\~ao Paulo, Brazil}
\email{oliveira@astro.iag.usp.br}

\author{Eduardo S. Cypriano}
\affil{Departamento de Astronomia, Instituto de Astronomia,
Geof\'{\i}sica e Ci\^encias Atmosf\'ericas da USP, Rua do Mat\~ao
1226, Cidade Universit\'aria, 05508-090, S\~ao Paulo, Brazil and\\
Department of Physics \& Astronomy, University College London,
London, WC1E 6BT} \email{cypriano@astro.iag.usp.br}

\author{Renato A. Dupke}
\affil{University of Michigan, Ann Arbor, MI 48109 and\\
Observat\'orio Nacional, Rua Gal. Jos\'e Cristino 77, S\~ao
Crist\'ov\~ao, CEP20921-400 Rio de Janeiro RJ, Brazil}
\email{rdupke@umich.edu}

\author{Laerte Sodr\'e Jr.}
\affil{Departamento de Astronomia, Instituto de Astronomia, Geof\'{\i}sica
e Ci\^encias Atmosf\'ericas da USP, Rua do Mat\~ao 1226, Cidade
Universit\'aria, 05508-090, S\~ao Paulo, Brazil}
\email{laerte@astro.iag.usp.br}


\begin{abstract}

Fossil groups are systems with one single central elliptical galaxy
and an unusual lack of luminous galaxies in the inner regions. The
standard explanation for the formation of these systems suggests
that the lack of bright galaxies is due to galactic cannibalism. In
this study we show the results of an optical and X-ray analysis of
\RXJ1340, the prototype fossil group. The data indicates that
\RXJ1340 is similar to clusters in almost every sense, dynamical
mass, X-ray luminosity, M/L and luminosity function, except for the
lack of L* galaxies.

There are claims in the literature that fossil systems have a lack
of small mass haloes, compared to predictions based on the LCDM
scenario. The observational data gathered on this and other fossil
groups so far offer no support to this idea.


%


Analysis of the SN Ia/SN II ejecta ratio in the inner and outer
regions shows a marginally significant central dominance of SN Ia
material. This suggests that either the merger which originated the
central galaxy was dry or the group has been formed at early epochs,
although better data are needed to confirm this result.


\end{abstract}


\section{Introduction}

Fossil groups (FGs) are bright and extended X-ray emission systems
(L$_{X,bol}$ $>$ 10$^{42}$ h$_{50}^{-2}$ ergs $s^{-1}$) dominated by
a single giant elliptical galaxy and with at least a two-magnitude
difference between the first and second-ranked galaxies (in the
R-band) within half of its virial radius. Initially, FGs were
thought to be the cannibalistic remains of small galaxy groups that
lost energy through tidal friction, perhaps the final stage of
compact groups (e.g. Barnes 1989, Ponman and Bertram 1993, Jones et
al. 2003). More recently, X-ray measurements of FGs were shown to be
inconsistent with this formation mechanism, in particular because
the intergalactic medium of a number of FGs is similar to those of
galaxy clusters, with temperatures sometimes in excess of 4 keV
\citep[e.g.][]{khosroshahi06a,khosroshahi07} and,  thus, FGs have
high gravitational masses compared with most nearby groups. As an
example, one of the most massive Hickson compact groups known, HCG
62, is an order of magnitude less massive than the typical FGs
studied so far \citep{claudia07}.

The high masses of FGs estimated in X-rays were confirmed by the
dynamical studies of these systems.  Recent measurements of galaxy
velocity dispersions in FGs
\citep{claudia06,cypriano06,khosroshahi07} are fully consistent with
the dynamical state of the system as determined from X-ray
observations, indicating that they have relatively deep
gravitational potential wells, typical of clusters.

The lack of L* galaxies in the central regions in a {\it
cluster-sized} potential was quite intriguing at first and motivated
a number of studies. \citet{donghia05} and \citet{Dariush07}
suggested that these systems are older than clusters of comparable
masses. von Benda-Beckmann et al. (2008), from the analysis of the
magnitude gap between the brightest and the second most bright group
member in a cosmological simulation, also suggest that fossil groups
are old structures and point out that they are a transitory phase in
the life of group which ends with the infall of new galaxies from
the neighborhood. \citet{diaz} analyzed FGs in the Millennium
simulation \citep{springel05} and in a mock catalogue and concluded
that although FGs have assembled early, first-ranked galaxies in FGs
only formed their magnitude gaps and their luminous galaxies quite
recently.

The most likely mechanism proposed for the lack of luminous galaxies
surrounding the central dominant galaxy is cannibalism
\citep[e.g.][]{jones03}.  The central merging, if not completely
``dry'' (i.e. involving gas-poor galaxies), should be accompanied by
star formation bursts, causing subsequent metal rich SN II-driven
galactic winds or superwinds \citep[e.g.][]{sanders96}.  These
secular winds would deposit metals and energy into the central gas
and this extra energy may contribute to explain the typical lack of
cooling cores in FGs \citep{sun04,khosroshahi04,khosroshahi07}. The
wind metal injection would make the central SN Ia/SN II ejecta of
FGs different (lower) from that of normal groups and similar sized
clusters. This scenario can, in principle, be tested by measuring
individual elemental abundances and their ratios with sufficiently
high S/N X-ray data \citep{dupke09}.

Based on the luminosity function of the fossil group \RXJ1340
published by \citet{jones00} and on the conclusion that \RXJ1340
lived in a sparse environment \citep{ponman94}, \citet{donghia04}
claimed that FGs may pose a severe problem for the cold dark matter
models, since they did not have as much substructure as expected for
such massive systems. On the other hand, \citet{sales07} using the
millennium simulation found that the luminosity functions of three
FGs, including \RXJ1340, are not inconsistent with $\Lambda$CDM
predictions, although the authors stress that the current sample of
FGs with reliable  luminosity functions is still poor. Reinforcing
this idea, \citet{zibetti}, who recently estimated the substructure
function from photometric data for six fossil systems (also
including \RXJ1340), finds that these systems are consistent with
normal clusters and with numerical simulations as well.  New
measurements and improvements in the determination of luminosity
functions for these systems is of paramount importance to determine
their origin. Here, we revise the luminosity function of the system
\RXJ1340, which is considered the prototype fossil group
\citep{ponman94}.

In Section 2 we describe the optical and X-ray data we used in this
work. In Section 3 we present the results we obtain using the
optical data. In Section 4 the results obtained from the X-ray data
are presented as well as the result of a test for the formation
scenario of galaxies through cannibalism of L* galaxies. Section 5
summarizes and discusses our findings. Throughout this paper, we
adopt, unless explicitly mentioned otherwise, a cosmology with
$H_{0}=70$ km s$^{-1}$  Mpc$^{-1}$, $\Omega_{0}=0.3$ and
$\Omega_{\Lambda}=0.7$, so that 1$^{\prime\prime}~\approx~$2.9 kpc
at the cluster distance ($z=0.172$, see Section 3.1).

\section{Observations and Data Reduction}

The imaging and multi-slit spectroscopic observations of the group
\RXJ1340 were done with the GMOS instrument, mounted on the Gemini North
telescope, on May 2nd and June 22nd, in 2006, respectively (see sections
2.1 and 2.2 below).  The X-ray study was based on archival Chandra data,
as described in section 2.3.

\subsection{Optical Imaging\label{sec:imaobs}}

\begin{figure}[h!]  \includegraphics[width=1.0\columnwidth]{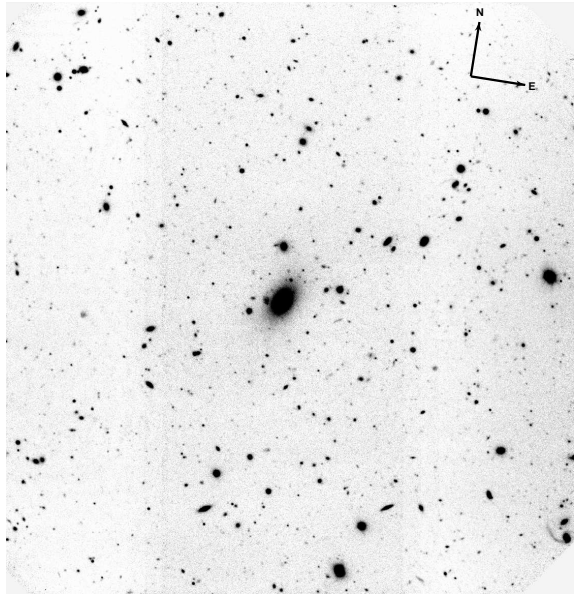}
\caption{\label{image} Optical i$^\prime$ image of \RXJ1340. The field
of view is 5.6 arcmin on a side, or $\sim$ 980 h$_{70}^{-1}$ kpc at the
object redshift.  The field orientation is shown in the upper right corner
of the figure.  } \end{figure}

The imaging consisted of $3\times200$s exposures in each of the two
filters from the SDSS system \citep{sloan} g$^\prime$ and i$^\prime$. The
typical FWHM for point sources was $\sim$ 0.75" in all images. All
observations were performed in photometric conditions.  Fig. \ref{image}
displays the i$^\prime$ image of the system.

The calibration to the standard SDSS system was made using
calibration stars observed about one month before the FG
observations. No calibration stars were taken in the night of the FG
observations, given that the telescope dome had to be closed due to
strong winds shortly after the object observation. We confirmed the
goodness of the i$^\prime$-band zero point obtained from the
calibration stars by comparing magnitudes measured by us with those
from the SDSS database, for 20 galaxies in common, in the magnitude
range i=18-20, finding no systematic difference. All observations
were processed in a standard way with the Gemini
IRAF\footnote[1]{IRAF is distributed by  NOAO, which is operated by
the Association  of Universities for Research in Astronomy Inc.,
under contract with the National Science  Foundation} package v1.8.

Positions and magnitudes (isophotal and aperture) were obtained for
all objects. We estimate that the galaxy catalog is essentially
complete down to  24.25 i$^\prime$ magnitude (the peak of the
number-count histogram). The SExtractor {\em stellarity} index
(Bertin and Arnouts 1996) was used to separate stars from galaxies.
All  objects with {\em stellarity} index $\le 0.8$ were selected as
galaxies.

\subsection{Optical Spectroscopy\label{sec:specobs}}

\begin{figure}[h!]  \includegraphics[width=1.0\columnwidth]{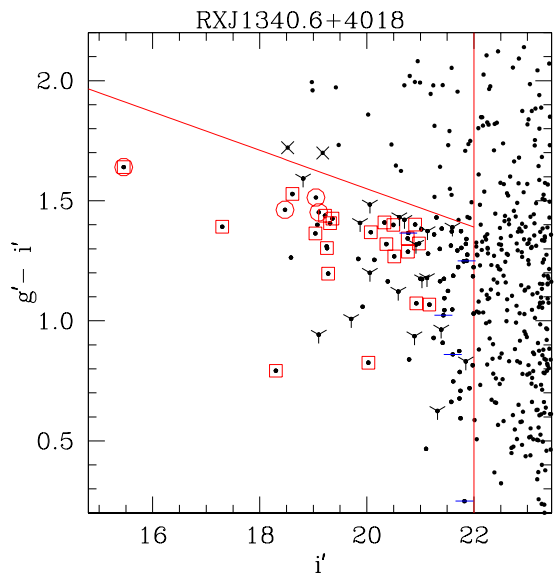}
\caption{\label{cmd} Color-magnitude diagram of the galaxies in the
\RXJ1340 field.  Points marked with squares (members), `Y' (non-members)
and  `--' (with spectra but no redshift) represent the galaxies observed
spectroscopically  by us with GMOS-N. The circles and `X' represent,
respectively, members and non-members published by  \citet{jones00}.
The inclined line indicates the upper limit for the cluster red-sequence
which we adopted when selecting the spectroscopic targets, whereas the
vertical line is the limit in magnitude up to which we chose objects
for spectroscopic follow up (i' = 22 mag).  } \end{figure}

Galaxies for spectroscopic follow-up were selected based on their
magnitudes and  colors. Figure \ref{cmd} shows the color-magnitude
diagrams for galaxies with \sloani$\le 23.5 $ mag. Objects in the
region within and bluer than the red cluster sequence  and brighter
than \sloani$=22$ mag ($M_{i'}<-17.6$, vertical line in Figure
\ref{cmd}) were selected as potential candidates. Galaxies above
the red sequence are expected to be in the background, since their
colors are redder than the expected colors of elliptical galaxies at
the group redshift. The outermost galaxy, which turned out to be a
member of the group/cluster, has a distance of 516 h$_{70}^{-1}$ kpc
from the X-ray center of \RXJ1340 (which coincides very well with
the central galaxy).

Four multi-slit exposures of 2400 seconds each was obtained through
a mask with 1.0\arcsec ~ slits, using the R400 grating, for a final
resolution of $\sim$7-8 \AA~(as measured from the FWHM of the arc
lines), covering approximately the range 4000 -- 8000 \AA~
(depending on the position of each slitlet).

Standard procedures were used to reduce the multi-slit spectra using
tasks within the Gemini {\sc IRAF} package. Wavelength calibration was
done using Cu-Ar comparison-lamp exposures before and after the exposures.

Redshifts for galaxies with absorption lines were determined using
the cross-correlation technique \citep{T&D} as implemented in the
package {\sc RVSAO} \citep{rvsao} running under {\sc IRAF}. The
final heliocentric velocities of the galaxies were obtained by
cross-correlation with several template spectra. The final errors on
the velocities were determined from the dispersion in the velocity
estimates using several different galaxy and star templates. In the
case of emission-line redshifts, errors were estimated from the
dispersions in redshifts obtained using different emission lines.
The S/N of the data,  measured in the continuum region
6400--6500\AA, ranged from 10 to 32.

Table \ref{tab_spectra} lists positions, isophotal magnitudes, aperture
(g' - i') colors, radial velocities with errors and the Tonry \& Davis
cross-correlation coefficient R for all galaxies with reliable velocity
determinations obtained in this study.

\begin{deluxetable}{lccccccc} \tablewidth{0pt} \tablecaption{Spectral
data for galaxies in the field of \RXJ1340 \label{tab_spectra}}
\tablehead{\colhead{(1)} & \colhead{(2)} & \colhead{(3)} & \colhead{(4)}
& \colhead{(5)} & \colhead{(6)} & \colhead{(7)}\\ \colhead{Name
\tablenotemark{a}} & \colhead{RA (2000)} & \colhead{DEC (2000)} &
\colhead{i$^\prime$ (AB Mag.)} & \colhead{g$^\prime$-i$^\prime$} &
\colhead{ $cz$ (km s$^{-1}$)} & \colhead{~R}} \startdata

G29.2+1704 & 13 40 29.2 & 40 17 04 & 20.98 & 1.32 & $~50379 \pm ~41$ & 4.1
           \\
G18.3+1830 & 13 40 18.3 & 40 18 30 & 19.26 & 1.30 & $~50455 \pm 137$ & 7.0
         \\
G30.7+1539 & 13 40 30.7 & 40 15 39 & 18.61 & 1.53 & $~50679 \pm ~41$ &
7.2 \tablenotemark{c}    \\ G26.4+1811 & 13 40 26.4 & 40 18 11 & 20.76 &
1.34 & $~50798 \pm ~57$ & 2.6
   \\
G35.9+1826 & 13 40 35.9 & 40 18 26 & 20.33 & 1.41 & $~50978 \pm ~39$ & 5.0
   \\
G44.2+1902 & 13 40 44.2 & 40 19 02 & 20.36 & 1.32 & $~51037 \pm ~60$ & 3.0
   \\
G23.6+1817 & 13 40 23.6 & 40 18 17 & 19.31 & 1.41 & $~51159 \pm ~48$ & 4.6
   \\
G32.5+1612 & 13 40 32.5 & 40 16 12 & 19.36 & 1.43 & $~51304 \pm ~46$ & 6.0
   \\
G44.5+1637 & 13 40 44.5 & 40 16 37 & 19.04 & 1.36 & $~51432 \pm 100$ & 3.0
 \\
G32.4+1533 & 13 40 32.4 & 40 15 33 & 20.77 & 1.29 & $~51476 \pm ~38$ & 5.6
   \\
G32.6+1910 & 13 40 32.6 & 40 19 10 & 19.22 & 1.44 & $~51566 \pm ~30$ & 7.8
   \\
G32.8+1740 & 13 40 32.8 & 40 17 40 & 15.46 & 1.60 & $~51577 \pm ~39$ & 8.9
   \\
G36.5+1843 & 13 40 36.5 & 40 18 43 & 20.50 & 1.40 & $~51718 \pm ~64$ & 4.0
   \\
G33.1+1753 & 13 40 33.1 & 40 17 53 & 20.90 & 1.40 & $~51769 \pm ~84$ & 2.4
   \\
G32.7+1934 & 13 40 32.7 & 40 19 34 & 20.92 & 1.07 & $~51777 \pm ~64$ & 3.0
   \\
G41.1+1558 & 13 40 41.1 & 40 15 58 & 20.07 & 1.37 & $~51926 \pm ~37$ & 6.2
   \\
G36.0+1604 & 13 40 36.0 & 40 16 04 & 20.51 & 1.27 & $~52004 \pm ~36$ & 5.9
   \\
G37.6+1517 & 13 40 37.6 & 40 15 17 & 17.30 & 1.39 & $~52034 \pm ~32$ & 7.2
   \\
G44.3+1612 & 13 40 44.3 & 40 16 12 & 19.28 & 1.20 & $~52362 \pm ~18$ &
\nodata\tablenotemark{b} \\ G40.4+1651 & 13 40 40.4 & 40 16 51 & 21.17 &
1.07 & $~52412 \pm ~96$ & \nodata\tablenotemark{b} \\ G20.4+1918 & 13 40
20.4 & 40 19 18 & 20.03 & 0.83 & $~52417 \pm ~52$ & 6.8 \tablenotemark{c}
\\ G20.3+1924 & 13 40 20.3 & 40 19 24 & 18.30 & 0.79 & $~52451 \pm ~27$
& 8.1 \tablenotemark{c}    \\
          &            &          &       &      &                  &
          \\
G24.0+1618 & 13 40 24.0 & 40 16 18 & 20.92 & 1.32 & $~58354 \pm ~85$ & 2.7
 \\
G23.9+1904 & 13 40 23.9 & 40 19 04 & 20.06 & 1.48 & $~60965 \pm ~39$ & 6.6
 \\
G24.8+1953 & 13 40 24.8 & 40 19 53 & 20.89 & 0.93 & $~61004 \pm 137$
& \nodata\tablenotemark{b} \\ G40.4+1909 & 13 40 40.4 & 40 19 09 &
18.81 & 1.59 & $~62707 \pm ~60$ & 4.5 \\ G18.3+1936 & 13 40 18.3 & 40
19 36 & 21.85 & 0.83 & $~64905 \pm ~84$ & \nodata\tablenotemark{b} \\
G27.1+1642 & 13 40 27.1 & 40 16 42 & 19.71 & 1.01 & $~66612 \pm ~44$ & 6.7
\tablenotemark{c}    \\ G21.5+1930 & 13 40 21.5 & 40 19 30 & 19.11 & 0.94
& $~85080 \pm ~56$ & 4.5 \tablenotemark{c}    \\ G40.8+1841 & 13 40 40.8
& 40 18 41 & 20.06 & 1.20 & $~90732 \pm ~53$ & \nodata\tablenotemark{b}
\\ G34.4+1906 & 13 40 34.4 & 40 19 06 & 21.32 & 0.62 & $108043 \pm
~18$ & \nodata\tablenotemark{b} \\ G40.3+1900 & 13 40 40.3 & 40 19
00 & 19.87 & 1.41 & $119768 \pm ~32$ & \nodata\tablenotemark{b} \\
G19.5+1700 & 13 40 19.5 & 40 17 00 & 21.12 & 1.18 & $135990 \pm ~16$
& \nodata\tablenotemark{b} \\ G42.5+1705 & 13 40 42.5 & 40 17 05
& 21.60 & 1.39 & $145587 \pm ~39$ & \nodata\tablenotemark{b} \\
G21.8+1959 & 13 40 21.8 & 40 19 59 & 21.39 & 0.96 & $155841 \pm
~19$ & \nodata\tablenotemark{b} \\ G27.7+2013 & 13 40 27.7 & 40
20 13 & 20.59 & 1.12 & $156092 \pm ~30$ & \nodata\tablenotemark{b}
\\ G19.9+1732 & 13 40 19.9 & 40 17 32 & 21.13 & 1.37 & $163895 \pm
~48$ & \nodata\tablenotemark{b} \\ G43.0+1701 & 13 40 43.0 & 40 17
01 & 21.03 & 1.17 & $181565 \pm ~54$ & \nodata\tablenotemark{b} \\
G26.6+1703 & 13 40 26.6 & 40 17 03 & 20.70 & 1.42 & $184426 \pm ~27$ &
\nodata\tablenotemark{b} \\ G37.8+1818 & 13 40 37.8 & 40 18 18 & 20.61 &
1.43 & $191963 \pm ~27$ & \nodata\tablenotemark{b} \\

\enddata \tablenotetext{a}{The names of the galaxies are based
on their 2000 celestial coordinates (RA seconds and DEC minutes
and seconds). Thus galaxy Gab.c+defg is located at 13 40 ab.c +40
de fg.} \tablenotetext{b}{Redshift measured from emission lines.}
\tablenotetext{c}{Spectra with emission lines but with redshift measured
from absorption lines.} \end{deluxetable}

\subsection{X-rays\label{sec:Xobs}}

We analized {\sl Chandra} archived data of the fossil group \RXJ1340 in
order to obtain its temperature and attempt a measurement of elemental
abundance ratios.  This system was observed by {\sl Chandra} ACIS-S3 in
Aug 2002 for 47.6 ksec. The cluster was centered on the S3 chip. The Ciao
3.3.0.1 with CALDB 3.2.4 was used, to screen the data.  After correcting
for a short flare-like period the resulting exposure time in our analysis
was 46.3 ksec. A gain map correction was applied together with PHA and
pixel randomization. ACIS particle background was cleaned as prescribed
for VFAINT mode. Point sources were extracted and the background used
in spectral fits was generated from blank-sky observations using the
{\sl acis\_bkgrnd\_lookup} script.

Here we show the results of spectral fittings with XSPEC V11.3.1 (Arnaud
1996) using the APEC and VAPEC thermal emission models. Metal abundances
are measured relative to the solar photospheric values of Anders \&
Grevesse (1989). Galactic photoelectric absorption was incorporated
using the WABS model (Morrison \& McCammon 1983). Spectral channels were
grouped to have at least 20 counts/channel. Energy ranges were restricted
to 0.5--8.0 keV. The spectral fitting errors are at the 1-$\sigma$ confidence
level,
unless stated otherwise. Given that individual abundances require a
higher S/N than that observed in the outer regions, we tied Oxygen and
Neon together since they have similar range of yield variation for SN Ia
and II.  Argon and Calcium are not well constrained and were also tied
together in the spectral fittings. Fits with these elements untied did
not produce  significantly different results.  For mass yields of SN Ia
and SN II, we use the values of Nomoto et al. (1997a,b). SN II yields
were calculated integrating over a Salpeter IMF for a progenitor mass
range of 10 to 50 \Msol.

\section{\label{res} Optical Properties of \RXJ1340}

\subsection{Galaxy velocity distribution}

\begin{figure} \includegraphics[width=1.0\columnwidth]{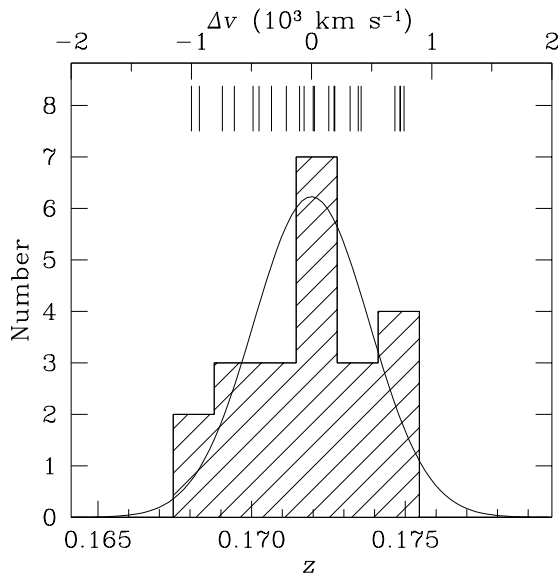}
\caption{\label{veldisp} Velocity histogram of \RXJ1340.  It shows
the distribution of the radial velocities of 22 galaxies in the inner
516  h$_{70}^{-1}$ kpc radius, with redshifts within $\pm$2000 km/s
of the systemic velocity of the group. The sticks on the upper part
of the plot show velocities of individual objects.  The {\sc rosat}
bi-weighted estimator gives a velocity dispersion  $\sigma = 565$ km/s and
a redshift of $\langle z \rangle = 0.1720$, for the sample of 22 objects.
} \end{figure}

Using the heliocentric radial velocities listed in Table
\ref{tab_spectra}, we consider as members of \RXJ1340 the 22
galaxies with velocities between 50378 and 52451 km s$^{-1}$ (out of
which only 5 have emission lines). The data were analyzed with the
statistical software {\sc rostat} \citep{beers}, which did not find
any large gap in the velocity distribution. In addition, no other
data-points were found outside a $\pm3\sigma$ range.  Fig.
\ref{veldisp} shows the velocity histogram for the 22 member
galaxies studied by us.

Using the robust bi-weighted estimator of {\sc rostat}, the
following values for the systemic redshift and velocity dispersion
were found: $\langle z \rangle = 0.1720\pm 0.0004 $ and $\sigma =
565 \pm 77 $ km s$^{-1}$, respectively. It is in contrast with the
earlier result of \citet{khosroshahi07} of $419\pm187$ determined
for only four galaxies, although not inconsistent, all uncertainties
considered. If we characterize the virial radius by the radius
within which the interior density is 200 times the critical density
(r$_{200}$), the measured projected velocity dispersion implies a
virial radius of r$_{200}=\sqrt{3}~\sigma/10~H(z)=1.29\pm0.18~Mpc$
(Carlberg et al. 1997).  The errors are due only to the velocity
dispersion determination and do not include the uncertainties
implicit in the cited equation (such as departures from a
$\rho~\propto~r^{-2}$ profile at large radii).  This value is
somewhat higher than that estimated from X-rays (see section 4),
which can be given, as defined in \citet{evrard96}, by r$_{200}\sim
0.88~\sqrt{(kT_{keV})}h_{70}^{-1}~Mpc = 1.00 \pm 0.05~Mpc$.

If the optically derived virial radius estimate is taken at face
value, it would imply that \RXJ1340 could not be classified
``strictly`` as a fossil group (in the Jones et al. definition)
given that there is one galaxy, G37.6+1517, which is 1.8 magnitudes
fainter than the brightest group galaxy in the i' band (1.6 in g'),
and which is located at a clustercentric radius of 480 kpc.  We note
that Santos et al. (2007) has also pointed out that \RXJ1340 did not
follow the strict definition of a fossil group.

We determine the dynamical mass of the system by using four
different mass estimators, as suggested by \citet{heisler85}:
virial, projected, average and median mass estimators (see results
in Table \ref{mass}). The adopted center of mass of the system was
the brightest cluster galaxy.  The errors were calculated by using
1000 bootstrap simulations. Note that the dispersion between the
mass values obtained using the several different methods is small
compared to the errors in a single measurement. The average of the
mass given by the four estimators is $3.4 \times 10^{14}$ M$\odot$.
We recall that all values are calculated within a radius of 516
h$_{70}^{-1}$ kpc ($\sim$ 40\% of the virial radius of the system).

\begin{deluxetable}{lcccc} \tablewidth{0pt} \tablecaption{Mass
Estimates \label{mass}} \tablehead{\colhead{(1)} & \colhead{(2)} &
\colhead{(3)} \\ \colhead{} & \multicolumn{2}{c}{} \\ \colhead{Estimator}
& \colhead{Mass ($10^{14}$ \Msol)} & \colhead{M/L$_{{g^\prime}\odot}$
(\Msol/L$_{{g^\prime}\odot}$) } \\} \startdata Virial     & $3.23\pm1.21$
& 330 & \\ Projected  & $4.34\pm1.42$ & 443 & \\ Average    &
$3.10\pm1.05$ & 316 & \\ Median     & $2.91\pm1.36$ & 297 & \\
          &               &     & \\
Mean value & 3.40          & 347 & \\ \enddata \end{deluxetable}

\subsection{The Luminosity Function and Mass-to-Light Ratios}

\begin{figure}[htq] \includegraphics[width=1.0\columnwidth]{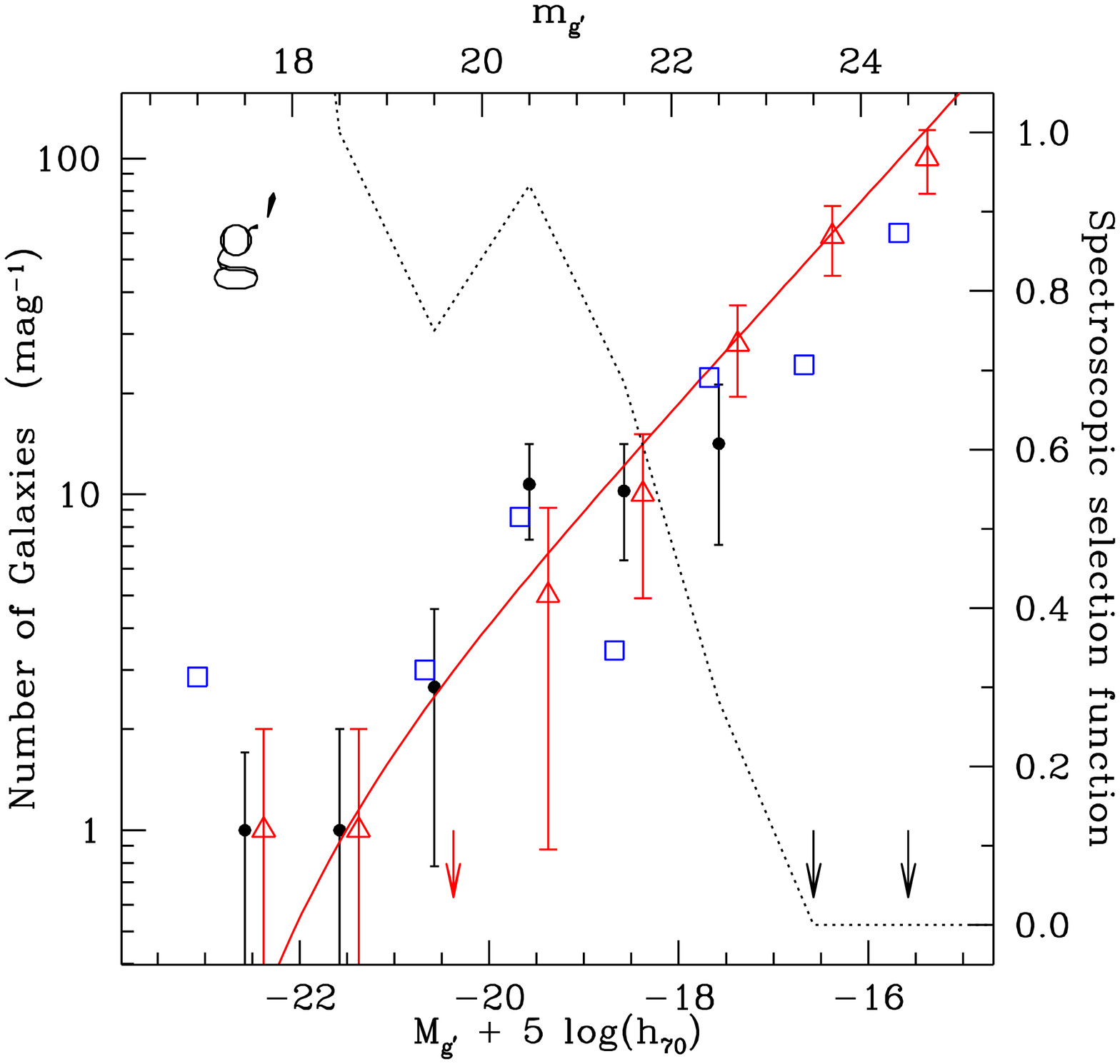}
\includegraphics[width=1.0\columnwidth]{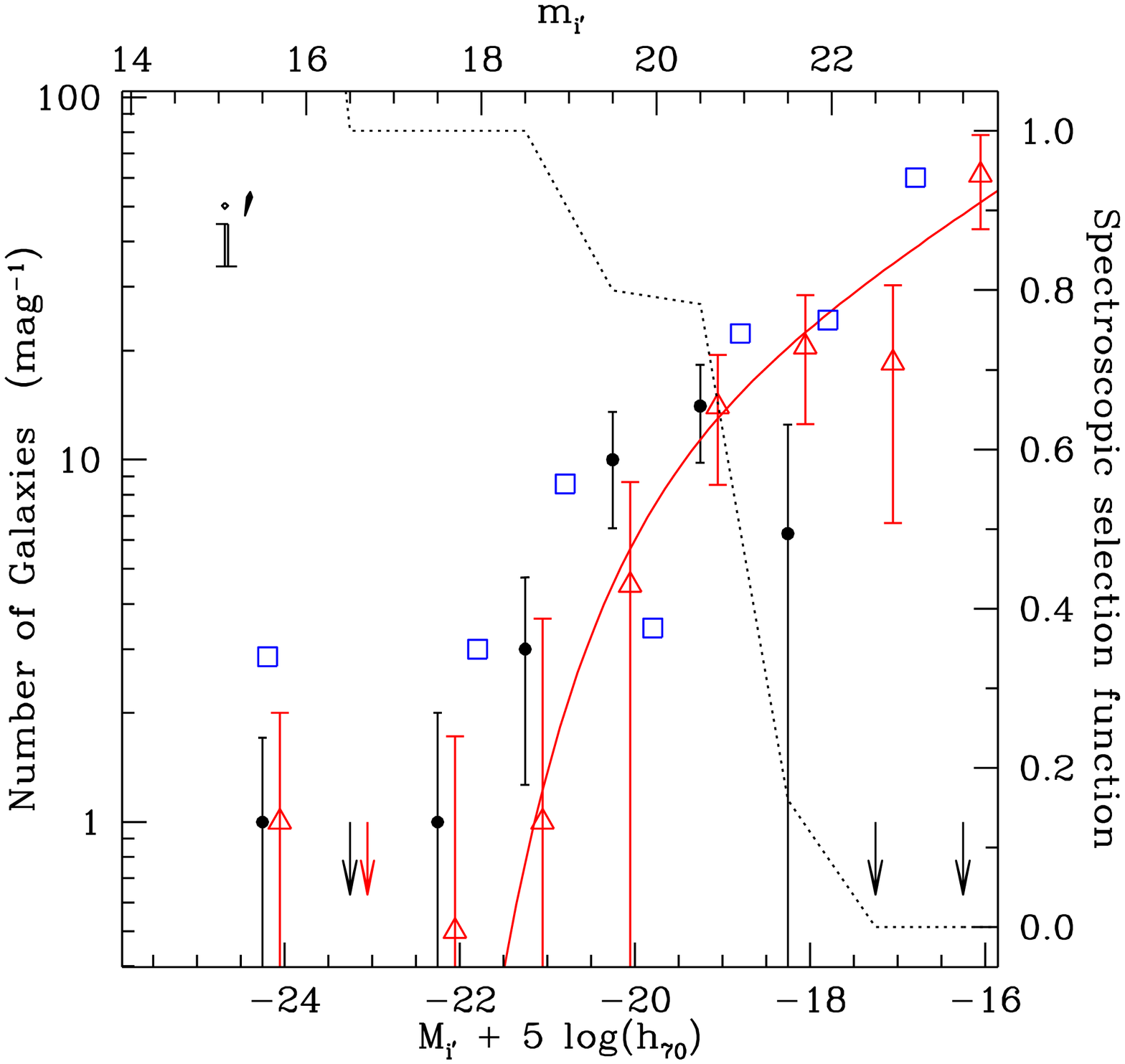} \caption{\label{flum}
Luminosity functions of \RXJ1340 in the g$^\prime$ and i$^\prime$
bands.  The solid circles show the completeness-corrected numbers of
spectroscopically confirmed members of \RXJ1340 per 1.0 magnitude bin
in the GMOS field. The error bars are 1$\sigma$ Poissonian errors. The
arrows show bins with number of galaxies less or equal to zero. The
dotted line is the selection function of the spectroscopic sample.
The open triangles show the photometrically-determined luminosity
function estimated through number counts and statistical subtraction
of the background. The points have been shifted by 0.2 mag (from the
center of the bin), for the sake of clarity.
The squares represent the luminosity function found for this group by \citet{jones00}.
The continuous lines
show the best fitted Schechter function of the photometric sample.
The brightest galaxy of the cluster was not included in the fit. The
agreement between the spectroscopically and photometrically-determined
luminosity functions is good, in the region of overlap.  } \end{figure}

We show in Fig. \ref{flum} the luminosity function of \RXJ1340 (solid
circles) for galaxies with spectroscopically confirmed membership either
obtained in this paper  (22 galaxies) or given by \citet[][three galaxies
plus the central one; for the latter we also have a spectrum]{jones03},
corrected for incompleteness. The absolute magnitudes were calculated
after correcting the observed magnitudes for Galactic extinction and
applying $k$-corrections. The selection function (number of galaxies with
reliable redshifts over the total number of galaxies in a given magnitude
bin), also shown in Fig.  \ref{flum}, was calculated considering only
galaxies bluer than the upper limit of the adopted red cluster sequence.

We have estimated the photometric luminosity function of \RXJ1340
down to the completeness level of our photometric data (g$^\prime$ =
24.5 and i$^\prime$ = 23.5 mag) by adopting the procedure described
in our previous papers \citep{claudia06, cypriano06}, where the
control fields used for the background subtraction are the ones from
\citet{natalia}. The photometric luminosity functions are shown in
Fig. \ref{flum} as open triangles. They go deeper than the
spectroscopic results (solid circles) and they suggest that the
number of galaxies keeps increasing at faint magnitudes. It is
important to note that, at bright magnitudes, there is no important
discrepancy between the luminosity functions calculated using both
methods.

In Figure \ref{flum} we also overplotted the data from \citet[][R$<$
400 h$_{50}^{-1}$ kpcR$<$ 400 h$_{50}^{-1}$ kpc]{jones00} for the
sake of comparison. For that we adopted the colours
g$^\prime$-R=0.90 and R-i$^\prime$=0.54, which implies an overall
g$^\prime$-i$^\prime$ colour of 1.44, which is consistent with our
data (see Figure \ref{cmd} and with the colours of an Sab galaxy at
the redshift of 0.17 \citep{fukugita}. By inspection of Figure
\ref{flum} one can see that Jones et al. and our data sets are
consistent with each other, although ours seem to be less noisy,
probably due to the use of colours to aid on the background
subtraction.

The shape of the luminosity function is very similar in the g' and i'-band
diagrams.  The faint end of the luminosity function is steeply rising
up to our completeness limit.  The shape of the photometric luminosity
function in the i'-band, when fit to a Schechter function in the interval
$-22.5 < M_{i^\prime}< -16.5$, is well described by such a function with
the following parameters: M$_{i^\prime}^* = -21.3 \pm 1.8$ and $\alpha =
-1.6 \pm 0.2$.  For the g'-band, the photometric luminosity function is
fit in the interval $-21.5 < M_{g^\prime} < -15.5$, and the best fit to a
Schechter function is given by the following parameters: M$_{g^\prime}^*
= -19.3 \pm 0.9$ and $\alpha = -1.6 \pm 0.2$. Lines indicating the best
Schechter luminosity functions fitted are plotted in Fig. \ref{flum}.

We integrated the best fit Schechter function on the \sloang band
to obtain the total luminosity of the group, assuming that the
magnitude of the Sun in the \sloang-band is M$_{{g^\prime}\odot}$ =
5.11 mag \footnote{Calculated by C. Willmer using the solar spectra:
www.ucolick.org/$\sim$cnaw/sun.html}. We then added the luminosity of the
central galaxy, since this galaxy is not taken into account when fitting
the luminosity function. The final result is 9.8 $\times$ 10$^{11}$ L\sol ~
on the \sloang band, where the central galaxy alone is responsible for
almost 80\% of the entire cluster luminosity budget. This leads to a
mass-to-light ratio of 347 M\sol/L$_{{g^\prime}\odot}$ (using the mean
value for the masses obtained with the four estimators,  $3.4 \times
10^{14}$ M$_\odot$ - see Table \ref{mass}).  Results for the mass-to-light
ratio, inside a radius of 516 h$_{70}^{-1}$ kpc ($\sim$ 40\% r$_{200}$),
using different mass estimators, are presented in Table \ref{mass}.

\subsection{Surface photometry of the BCG}

\begin{figure}[!ht] \includegraphics[width=\columnwidth]{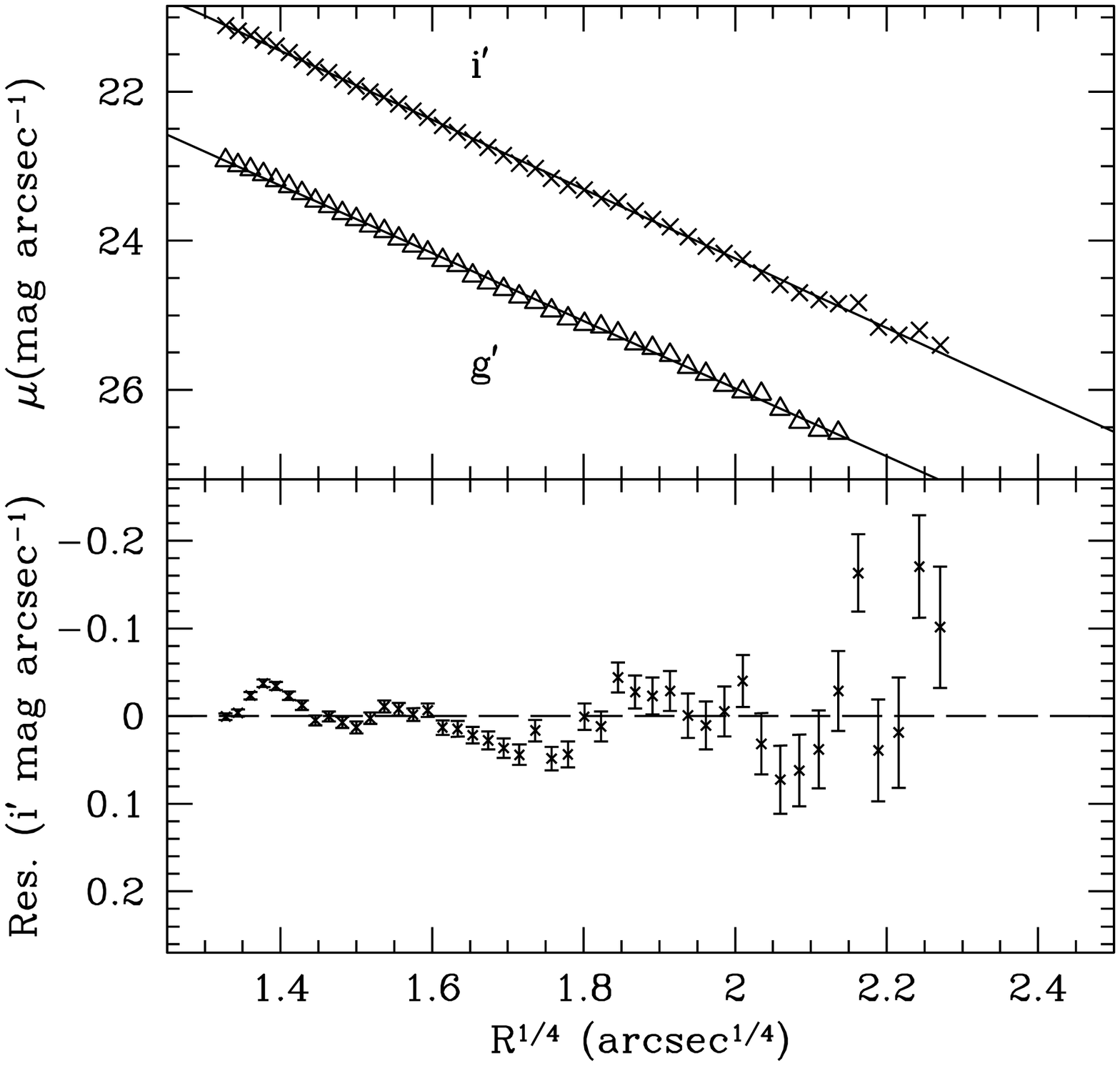}
\caption{\label{phot} {\it Upper panel:} Photometric profile of the
central galaxy of \RXJ1340. We show the isophotal levels as a function
of the semi-major axis to the power 1/4. The solid line is the best
fit to the de Vaucoleurs profile in the region where r$>$3.0 arcsec and
$\mu_{i^{\prime}}=25.0$ mag arcsec$^{-2}$. {\it Lower panel:} residual
between the actual r$^\prime$-band profile and the de Vaucoleurs profile
fit, showing no significant excess that would characterize a cD galaxy.}
\end{figure}

In the upper panel of Fig. \ref{phot}, the azimutally averaged photometric
profile of the central galaxy of \RXJ1340 is shown. The surface photometry
was performed using the task {\sc ELLIPSE} in {\sc STSDAS/IRAF}, which
fits ellipses to extended object isophotes.  We allowed the ellipticity
and position angle of the successive ellipses to change but the center
remained fixed.  The ellipse fitting was performed only in the deeper
i$^\prime$ image. For the g$^\prime$-band image, the software measured the
isophotal levels using the parameters estimated in the i$^\prime$-band
image.  There are several small objects within the isophotes of the
central galaxy which were masked during the profile fitting procedure.

We have fitted a r$^{1/4}$-law to the galaxy profile, from well
outside the seeing disk (3.0\arcsec) to a radius corresponding to
$\mu_{i^{\prime}}=25.0$ mag arcsec$^{-2}$.  In the lower panel of
Fig. \ref{phot}, the residuals (data -- r$^{1/4}$-law model) for the
i$^\prime$-band data are shown. There is no light excess over the
de-Vaucoleurs profile indicating that this galaxy is not a cD.

\section{X-ray properties of \RXJ1340}

\subsection {ICM temperature and abundances}

The gas temperature is consistent with a flat profile with a central
(r$<$37$^{\prime\prime}$) value of 1.21$\pm$0.07 keV and outer value
(112$^{\prime\prime}>$r$>$37$^{\prime\prime}$) of 1.31$\pm$0.14 keV,
consistent within the errors with the  average value of 1.16$\pm$0.08
keV determined by \citet{khosroshahi07}. If a cold core is present,
it is very mild and not seen with the current statistics. The iron
abundance is  consistent with a flat profile, but shows marginal signs
of a central enhancement.  Its central value is 0.29$\pm$0.09 Solar
and in the outer regions we can only set an upper limit of 0.2 Solar.
The measured value of the intracluster gas X-ray luminosity from 0.5--10
keV of \RXJ1340 is $\sim 2.3\times10^{43}$ h$_{50}^{-2}$ erg\,s$^{-1}$.
According to the relations for groups and clusters from \citet{mahdavi01},
we find that for a system with this X-ray luminosity we would expect
a velocity dispersion of about $\sigma$ = 425 km s$^{-1}$, in good
agreement with our direct measurement, given the high scattering in the
L$_{X}$--$\sigma$ relation ($\pm$240 km/s).


\subsection {Central abundance ratios and fossil group formation models}

The isolation of the central galaxy in FGs and the correlation between the FG's X-ray and central galaxy's optical luminosities   suggest that the formation process
of these systems involve the merging of central bright galaxies due to dynamical
 friction. Given the time scales for dynamical friction, it was originally suggested that FGs have an earlier formation epoch than regular groups (e.g. Ponman et al. 1994). This is also in agreement with predictions of the concentration parameters and magnitude differences of the two brightest galaxies in FGs from numerical simulations  (Wechsler et al. 2002, D'onghia et al. 2005). However, the cooling time of FGs is observed to be significantly below
the Hubble time (e.g., RX J1416.4+2315, ESO 3060170, Sun et al. (2004); NGC 6482,
Khosroshahi et al. (2004, 2006), but they typically \textit{lack cooling cores}, in disagreement with what one would expect for an old undisturbed system. This is a notable difference from regular rich groups, which often show cooling cores (e.g. Finoguenov \& Ponman 1999), even in spite of strong AGN activity.

To try to solve this apparent contradiction it is desirable to have
other independent formation age indicators. Elemental abundances
measured in the central regions of FGs can be used as such
\citep{dupke09}. The general idea is based on the fact that
different metal enrichment mechanisms contaminate the intragroup
medium with  different relative ejecta dominance and are more
intense at different cluster locations and in different time scales.
For example, protogalactic winds are dominated mostly by SN II and
produce an early more distributed background of SN II ejecta
throughout the group/cluster, while ram-pressure stripping happens
continuously throughout the cluster's history and creates a more
concentrated SN Ia contamination, given its dependence on the
ambient density.

If the group has an earlier origin and has been undisturbed for a
long time, the above mentioned enrichment processes will develop a
radial chemical gradient in such a way as to increase the central Fe
mass fraction dominance from SN Ia, which is observed in rich groups
and clusters of galaxies (e.g. Dupke 1999; Dupke \& White 2000a,b;
Allen et al 2001; Finoguenov et al. 2000; Dupke \& Arnaud 2001). If
the mergers are "dry", i.e., involving gasless galaxies, an older
system should have a higher SN Ia central enhancement than a younger
system. This central SN Ia enhancement has not been observed yet in
any FG. In fact, most FGs seem to have an inverted trend of reduced
central SN Ia dominance \citep{dupke09}, suggesting that, without
fine tuning, the difference between FGs and regular groups is not
due to age but to the type of merger that created the central
galaxy. Gas rich galaxy mergers presumably would be accompanied by
SN\ II\ powered winds, diluting the SN Ia central dominance.

In order to compare the enrichment distribution, we selected two
regions for spectral analysis. The first corresponds to the typical
cooling radius, where the cooling time is less than the Hubble time
(assumed as $\sim$ 100kpc or 10\% of r$_{200}$, which is  determined
according to r$_{200}\sim 0.88~\sqrt(kT_{keV})h_{70}^{-1}$~Mpc) and
is denoted r$_{cool}$. The second region is an annulus going from
the border of r$_{cool}$ to $\sim$ 320 kpc, denominated outer in
Table 3.  A central circular region with 5~kpc radius was excluded
to account for possible AGN contamination.  Even though the current
observation does not allow us to determine variations of the
abundance of different $\alpha$-elements significantly, there seems
to be an overall tendency for the abundance of the
$\alpha$--elements silicon, sulfur, oxygen and magnesium to decline
towards the central regions. The results are shown in Table 3 and
suggest a gradient in the supernovae type enrichment dominance,
unlike that of other FGs \citep{dupke09}, but similar to that found
in rich groups and poor clusters. The number of counts is very small
for the analysis of individual ratios, so that we can only analyze
an ensemble of abundance ratios. The error weighted average of the
abundance ratios O/Fe, Si/Fe, S/Fe and Mg/Fe indicates that the SN
Ia Fe mass fraction in the central regions of \RXJ1340 is 91$\pm$9\%
and consistent with $\sim$0\% in the outer parts.

\begin{deluxetable}{lccc} \small \tablewidth{0pt} \tablecaption{Individual
Elemental Abundance Profiles\tablenotemark{a}} \tablehead{
\colhead{Element} & \colhead{Inner r$_{cool}$\tablenotemark{b}}
& \colhead{Outer\tablenotemark{b}} } \startdata O &
0.0$^{+0.15}_{-0.0}$\tablenotemark{c} & 0.4$^{+2.0}_{-0.40}$
\nl Mg & 0.41$^{+0.46}_{-0.36}$ & 1.55$^{+1.95}_{-0.85}$
\nl Si & 0.26$^{+0.29}_{-0.22}$ & 0.44$^{+0.96}_{-0.34}$
\nl S & 0.0$^{+0.26}_{-0.0}$ & 1.78$^{+3.72}_{-1.78}$  \nl
Fe & 0.29$^{+0.10}_{-0.09}$ & 0.0$^{+0.2}_{-0.0}$ \enddata
\tablenotetext{a}{Abundances are measured relative to the solar
photospheric values of Anders \& Grevesse (1989), in which Fe/H = 4.68
$\times$ 10$^{-5}$ by number} \tablenotetext{b}{$\chi_{\nu}^{2}$ for
the spectral fittings are 1.16 and 1.52 for 39 and 110 d.o.f. for the
inner and outer regions respectively} \tablenotetext{c}{Errors are 68\%
confidence limits} \end{deluxetable}
 \clearpage

\section{Discussion}

Here we summarize our main findings in this optical and X-ray study
of \RXJ1340:

\begin{itemize}

\item
\RXJ1340 is a cluster-like fossil group at $z=0.172$, with
a velocity dispersion of 565 km/s and a mass of  $\sim$3.4 $\times$
10$^{14}$ \Msol within 516 h$_{70}^{-1}$ kpc (about 40\% of its virial
radius). Thus, the prototype FG is indeed a cluster, not a group.

\item We find a steep faint-end for the galaxy luminosity function
of the cluster (with $\alpha = -1.6 \pm 0.2$),  which is
consistent with the data points published by Jones et al. (2000)
and also with the Virgo and Coma luminosity functions. We
conclude that there is little room for claims that the low mass
(faint) end of fossil groups is essentially different from that
of clusters such as Virgo \citep{donghia04}. Thus, they do not
pose a problem for CDM models \citep[see also][]{zibetti}.

\item
The brightest object of \RXJ 1340 is not a cD galaxy. Its
surface brightness profile follows closely a de Vaucouleurs-law
out to 2.5 effective radii.

\item The X-ray analysis of the elemental abundance ratio profiles
indicates that the Fe mass fraction in the central region is
dominated by SN Ia contamination, similar to other groups and
poor clusters of galaxies. This suggests that either the merger
which originated the central galaxy was ``dry'' (the galaxies
involved in this process were gas poor), or the the group has
been formed at early epochs.

\item
\RXJ1340 does not constitute a low-density environment. Like \1552
and RX~J1416.4+2315, this object is a galaxy cluster with a large
magnitude gap in the bright end.  The fairly high X-ray emission, the
large fraction of elliptical galaxies (most of the bright galaxies in
Fig. 1 are early-types), the radial velocity distribution (Fig. 3),
as well as the lack of obvious substructures, imply a high degree of
virialization for \RXJ1340.

\end{itemize}

Fossil groups have, by definition, a lack of bright galaxies because
of the selection criteria used to catalogue them. The bright-end of the
luminosity function of these systems is then known to be unusual, with
too few L$^*$ galaxies.  At the faint-end, not much has been known so
far, but a scenario seems to be emerging that the massive FGs have steep
luminosity functions with exception of perhaps RX~J1552.2+2013, for which
we just reach the magnitude of the dwarf upturn (the point where the curve
goes from being giant-dominated to dwarf-dominated). For the particular
case studied here, of \RXJ1340, the faint end of its luminosity function
is very steep ($\alpha = -1.6 \pm 0.2$) but, within the errors, it
is similar to that of other galaxy clusters of comparable masses. For
 example, for the 2dF and RASS--SDSS clusters, $\alpha \simeq -1.3$
in the blue-band \citep{propris,popessoII}. It is also comparable with
the faint-end slope of clusters like Virgo
\citep[$\alpha = -1.28 \pm 0.06$,][]{rines08}
and Coma \citep[$\alpha = -1.47 \pm 0.09$,][]{iglesias03}.
It is worth mentioning that the above errors are statistical and, due to
cosmic variance, they are probably underestimated.

We found that the central galaxy of \RXJ 1340 does not show any
light excess over its de Vaucouleurs profile (which would
characterize a cD galaxy). So far, only one central galaxy of a FG,
that of RX~J1552.2+2013, has been classified as a cD (perhaps not
coincidentally, it is also the only FG observed to have a declining
luminosity function at $\sim$M$^*+4$).

The best accepted scenario for the formation of FGs involves the
merging of the central bright galaxies. The lack of SN II ejecta
dominance in the central regions of \RXJ1340 implies that the
mergers that formed the central galaxy were "dry". The overall
results are consistent with an early epoch origin for this FG, so
that it has been undisturbed for a long  time and central
replenishment with SN Ia material, through e.g., ram-pressure
stripping, has not been affected by secondary SN\ II winds.  The
suspected lack of a cooling core in this system is, if this result
is confirmed with better data, still an open question. The scenario
of fossil group formation may become clearer when more of these
groups are studied spectroscopically. Ongoing determinations of the
luminosity function for a large sample of fossil groups and the
detailed study of the properties of the brightest group members,
including the determination of their ages and metal abundances, may
also elucidate some of the unsolved problems regarding these
systems.


\begin{acknowledgements}

We are grateful to the referee for comments that helped improving
the paper and to H. Khosroshahi for communicating velocities for
three members of the group, which were used in the determination of
the spectroscopic luminosity function.  We thank Raimundo Lopes de
Oliveira for a careful reading of the manuscript and useful
suggestions.  We would also like to thank the Gemini staff for
obtaining the observations. The authors would like to acknowledge
support from the Brazilian agencies FAPESP (projeto tem\'atico
06/56213-9), CNPq and CAPES. Dupke also acknowledges support from
NASA Grants NNX07AH55G, NAG 5-3247, GO4-5145X, GO5-6139X,
NNX06AG23G, NNX07AQ76G, NNX08AB70G \& FAPESP grant 06/05787-5. We
made use of the Hyperleda database and the NASA/IPAC Extragalactic
Database (NED). The latter is operated by the Jet Propulsion
Laboratory, California Institute of Technology, under contract with
NASA.

\end{acknowledgements}




\end{document}